# Resource CoAllocation for Scheduling Tasks with Dependencies, in Grid

Diana Moise [1,2], Izabela Moise [1,2], Florin Pop [1], Valentin Cristea [1]

[1] *University "Politehnica" of Bucharest, Romania*
[2] *INRIA/IRISA, Campus de Beaulieu, 35042 Rennes Cedex, France*
{diana.moise, izabela.moise}@irisa.fr, {valentin, florin.pop}@cs.pub.ro

## Abstract

*Scheduling applications on wide-area distributed systems is useful for obtaining quick and reliable results in an efficient manner. Optimized scheduling algorithms are fundamentally important in order to achieve optimized resources utilization. The existing and potential applications include many fields of activity like satellite image processing and medicine. The paper proposes a scheduling algorithm for tasks with dependencies in Grid environments. CoAllocation represents a strategy that provides a schedule for task with dependencies, having as main purpose the efficiency of the schedule, in terms of load balancing and minimum time for the execution of the tasks.*

## 1. Introduction

A Grid application usually requires the coordinated processing of complex workflows, including scheduling of heterogeneous resources (computational, storage, network etc) distributed over different administrative domains [1]. The Grid scheduler is required to plan and coordinate the workflow execution, which involves: required resources reservation, developing a complete workflow in advance, cost management and accounting.

Over the past years, the researches in the domain have been directed towards finding solutions to the key problem of the *efficiency of the scheduling*. An important method for improving the efficiency of the scheduling is that of ensuring that the provided schedule has the property of *load balancing* of the system resources. Load balancing refers to a balanced assignment of tasks to resources, without overloading a resource or overusing it, in comparison to the others. Load balancing represents the process of maintaining balanced workloads across multiple CPU's or systems. By scheduling tasks in a balanced manner, the completion time of the tasks is reduced and the number of resources used simultaneously is increased.

Grid scheduling involves more than an efficient algorithm: data storage mechanisms, data access mechanisms, data sharing mechanisms, the security of the environment the data is shared in, communication protocols between the sharing entities. Scheduling algorithms can be divided into two major categories: *static* and *dynamic* [2]. *Static Scheduling* assumes that the information regarding all the resources and tasks required by an application is available at the scheduling moment. The placement of an application is static; the main advantage drawn from this approach is that computation cost can be estimated before the actual execution. On the other hand, it is difficult and in some cases impossible, to know *a priori* the state of the resources and the specifications for the tasks that must be scheduled. In *Dynamic Scheduling*, task allocation is performed during the actual execution of an application; it can be applied when it is difficult to estimate the cost of an application (execution time, direction of branches), or when jobs arrive for execution in a real-time mode.

Another classification of scheduling algorithms consists in *centralized* and *decentralized*. In the case of a *centralized* algorithm, global scheduling decisions are made by a central scheduler, which is also the single access point to the whole infrastructure; the central scheduler manages the other scheduling instances (that interact with the local resources). A *decentralized* algorithm assumes that scheduling decisions are made by all the scheduling instances; each of them can accept tasks for execution. A decentralized strategy has the advantage of a greater reliability, because it is less subject to single points of failure in the system, than a centralized mechanism. The well functioning of the system is evenly distributed between all the entities of the system, therefore, if an entity fails, the system can continue to function. In the case of a centralized algorithm, if the entity that fails is the centralized one, the system will no longer be able to function.

This paper proposes the CoAllocation strategy which is a decentralized, dynamic and optimal mechanism for task scheduling in Grid environments. The paper is structured as follows. Section 2 provides a general presentation of the concepts the strategy involves. Section 3 describes the structure of the system CoAllocation runs on, the entities and the protocols involved in the mechanism. Section 4 refers to the testing and the evaluating methods by which the strategy's performances can be assessed. Section 5 refers to related work. Section 6 provides a general overview of the main ideas presented in the previous chapters.

## 2. CoAllocation – a general presentation

In general, CoAllocation ensures that a given set of resources is available for use, simultaneously. Tasks with dependencies refer to the fact that a task requires, in order to execute the results obtained by executing another task; in other words, the output of a task's execution is part of the input of a task that depends on it. Tasks with dependencies are parts of a set of action which consist an activity. CoAllocation is a strategy which schedules tasks with dependencies, having as main purpose, the efficiency of the schedule, in terms of load balancing and minimum time execution of tasks.

Therefore, CoAllocation is a method that receives as input a set of tasks with dependencies and provides as output a schedule which maps tasks to resources, a resource being mapped to several tasks. A condition the schedule must fulfill is that all the tasks must finish their execution before a *deadline;* the schedule must be an efficient one, in terms of execution time of the tasks.

### 2.1. Description of a task

A task represents a specific piece of work required to be done as part of a job or application. In CoAllocation strategy, a task is described by the following parameters: *taskId* which represents a unique identifier for a specific task; *processingTime* specifies the amount of time the task requires for execution; *requirements* describe the parameters a task requires for execution: Memory – specifies how much memory the tasks needs; cpuPower – specifies how much cpu power the task needs; deadlineTime – specifies the exact moment of time the task must finish its execution; *dependencies* refer to the tasks that a specific task depends on; for each task, there must be specified the task identifier and a communication time, which represents the necessary amount of time (expressed in seconds) required by the transfer of data, code etc, between a task and the task that depends on it;

A general specification for a task is presented below:

```
<task>
    <taskId>…</taskId>
    <requirements>
        <memory>…</memory>
        <cpuPower>…</cpuPower>
        <deadlineTime>…
        </deadlineTime>
    </requirements>
    <processingTime>…
    </processingTime>
    <depends>
        <taskId>…</taskId>
        <commTime>…</commTime>
    </depends>
</task>
```

The specifications for several tasks are contained in XML files.

### 2.2. Description of a resource

A resource or a node consists of one or more central processing units (CPU's).

A specification of a resource, in Advance Reservation, consists of the following: *Id* represents a unique identifier by which the node can be referred to; *NodeName* specifies the name of the node; *ClusterName* specifies the name of the cluster the node belongs to; *FarmName* specifies the name of the farm the cluster belongs to; *Parameters* specify the characteristics of the processing resource: CPUPower, Memory, CPU_idle.

A specification for a resource has the following general format:

```
<Node>
    <Id>…</Id>
    <FarmName>…</FarmName>
    <ClusterName>…</ClusterName>
    <NodeName>…</NodeName>
    <Parameters>
        <CPUPower>…</CPUPower>
        <Memory>…</Memory>
        <CPU_idle>…</CPU_idle>
    </Parameters>
</Node>
```

### 2.3. DAG

A set of tasks with dependencies can be represented as a Directed Acyclic Graph [4]. A task represents a node in the DAG; a dependency between two tasks represents the directed edge that connects the corresponding two nodes which contain the tasks. This type of DAG has costs both on an edge and inside a node:

· Inside a node, the cost represents the execution time of the task;

· Across an edge, the cost represents the amount of time required by the transfer of data, code, files etc, between the two tasks inside the two nodes the edge connects;

# 3. The functioning mechanism

## 3.1. System anatomy

CoAllocation is based on a system with a structure composed by the following entities:

· *A broker* represents an entity designated to provide the interface with the user. Its role within the algorithm is different from that of an agent. The broker receives a set of tasks from the user and provides, in the end, a schedule of the given set on the available resources. The broker does not maintain state information about the resources of the system. A broker communicates with the agents in the system.

· *An agent*'s main role is to maintain state information about the resources it is responsible for. A resource consists in one or more CPU's. CoAllocation is a decentralized method, which means that every agent keeps information about the local resources it manages and it is responsible for them. (as opposed to a centralized strategy which requires the presence of an entity which manages all the resources of the system). Another important function of an agent is to communicate with the broker.

Due to the decentralized strategy that CoAllocation uses, the method is more reliable than a centralized one, being less subject to single point faults.

Figure 1 shows the entities of the system, and the communication between them:

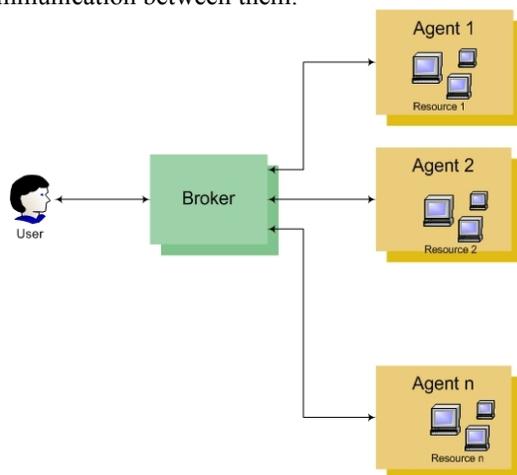

**Figure 1. The entities of the system**

## 3.2. The algorithm

CoAllocation uses an algorithm that consists in three phases:

### 1. Task clustering

The first phase of the algorithm receives as input a set of tasks and provides as output a set of clusters. The broker receives the name of an XML file from the user, builds the set of tasks the user wants to schedule and then initiates the clustering algorithm. A cluster is a set of tasks, representing a part of the initial DAG. The broker is responsible for building the clusters so that the graph obtained by replacing a set of tasks with the corresponding cluster will also be a DAG.

The clustering algorithm has the following steps:

- initially, each node is a cluster;

- every step combines the clusters from the precedent step, based on the dependencies between the combined clusters; thus, clusters C1 and C2 are merged if there are any dependencies between tasks belonging to cluster C1 and tasks form cluster C2.

- a condition every step must check and accomplish is that the graph formed by the clusters does not have any cycles; if that happens, the cluster which leads to a cycle, is not combined with another cluster;

In order to obtain a schedule with the property of load balancing, the clusters should contain almost the same number of tasks. The broker tries to build each cluster based on task dependencies, so that it contains the maximum number of tasks (which is equal to number_of_tasks / number_of_agents + 1). Thus, when building a cluster from two clusters at the current step, the broker looks for dependencies between tasks belonging to the two clusters: if C is the current cluster, the broker searches for a cluster that contains tasks that depend on tasks from cluster C; if the merge is possible (there are no cycles in the newly formed DAG and the new cluster does not exceed the maximum number of tasks), the next step will look for a cluster to merge with the one formed at the previous step; this mechanism is applied until one of the following conditions is fulfilled: there is no cluster that can be merged with the current cluster, the number of tasks in the cluster is greater than number_of_tasks / number_of_agents + 1. This strategy ensures that clusters are built in a maximal manner, based on tasks dependencies and accomplishing the condition regarding the number of tasks each cluster contains.

The equivalent graph, which has clusters as nodes, has edges which connect the clusters, corresponding to the initial DAG. In order to keep the DAG concept, an edge in the DAG of clusters, has as maximum value for its cost, the sum of the costs of all the edges from the initial DAG which connect the nodes from the two clusters.

## 2. Dynamic scheduling (inside a cluster)

The second phase begins with the distribution of each cluster to a corresponding agent. The broker sends each cluster to an agent that must schedule the cluster's tasks on its local resources. While distributing the clusters, the broker computes the number of tasks sent to each agent and always sends the current cluster to the agent that received the minimum number of tasks. Therefore, the initial set of tasks is evenly distributed between the agents.

All the agents work simultaneously, scheduling the tasks based on the dependencies between them and ignoring the dependencies between clusters. Therefore, each agent will begin the scheduling phase starting with time 0. As the local scheduling is done simultaneously by all the agents, the overall scheduling time is considerably reduced.

The dynamic scheduling inside a cluster uses a version of the *hybrid remapper algorithm* [3].
This algorithm has two steps:
- Level decomposition: the set of tasks is divided into blocks, so that there are no dependencies between the tasks of each block; the first block contains tasks that have no dependencies; the nth block groups all the tasks which depend on the tasks of the precedent blocks (from 1 to n − 1);
- The scheduling algorithm consists of:

```
For-each block k, 0 < k <= n
    For-each task t in block k
        schedule t on each
            resource, based on
            the start time of the
            tasks it depends on
        choose the resource on
            which t can be
            scheduled at the
            earliest time
    End-for
End-for
```

The algorithm has a dynamic nature because the tasks cannot be scheduled prior to the algorithm; the resources change in time, a task scheduled at each step depends on the tasks scheduled at the precedent steps.

## 3. Dynamic Scheduling of the clusters

The last phase of the algorithm schedules the graph of clusters, based on the dependencies between the clusters. The broker decomposes the cluster DAG into levels; then, it receives replies from the agents that scheduled the clusters of the first level and builds the final schedule, maintaining for each task the node it was scheduled on, start time and end time; for the next levels, the broker sends to the agent that scheduled the cluster, the (end time + communication time) for each task the tasks of the cluster depend on; the agent delays the start time of each scheduled task according to the information received from the broker, sending a reply with the scheduling it has made.
The diagram shown in figure 2 summarizes the phases of the algorithm:

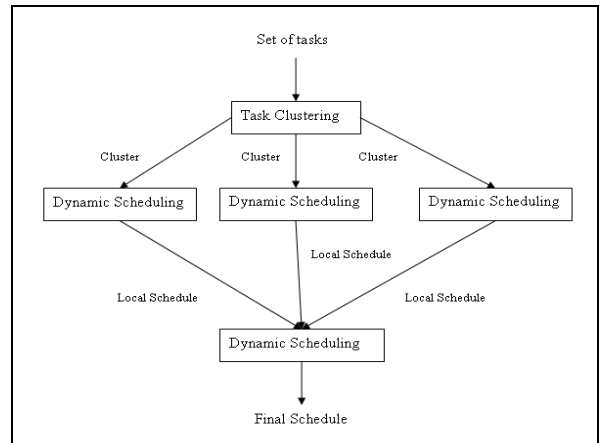

**Figure 2. Algorithm workflow**

## 3.3. Communication Protocol

The algorithm presented in the previous paragraph involves three communication protocols between the entities of the systems:

*User – Broker Communication*

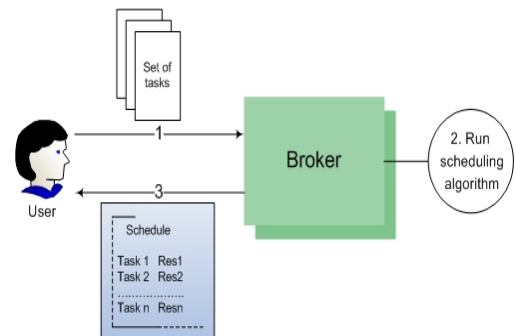

**Figure 3. Protocol 1**

The user types a file name as input to the broker, file that contains the description of a set of tasks. The broker parses the XML file and starts the scheduling algorithm for the set of tasks. The result of the algorithm is a set of mappings (taskId, resourceId) that specifies the name of the resource on which each task was scheduled. This set of mappings is returned to the user, as the schedule for the initial set of tasks.

*Broker – Agent Communication (1)*

Figure 4 shows the communication protocol between the broker and an agent that received a cluster of tasks, cluster that is situated on the first level of the level decomposition the broker produced. The broker sends the cluster of tasks to the agent, the agent then runs the local scheduling algorithm on its resource and replies to the broker, sending the schedule it obtained. The schedule is used by the broker to build the final schedule.

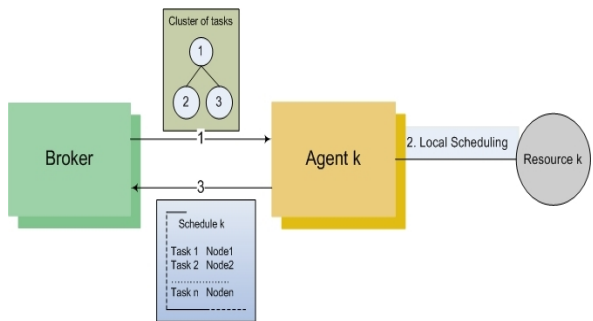

**Figure 4. Protocol 2**

*Broker – Agent Communication (2)*

Figure 5 describes the communication steps and the information exchanged between the broker and an agent that is assigned a cluster that doesn't belong to the first level in level decomposition of the DAG of clusters. In addition to steps 1, 2 and 5 that are the same as steps 1, 2 and 3 from the previous type of communication, the protocol contains steps:
3 – the broker sends to the agent, scheduling information about the tasks that the tasks in the cluster depend on (taskId, (endTime + commTime));
4 – the agent modifies the schedule obtained at step 2, according to the information received at the previous step.

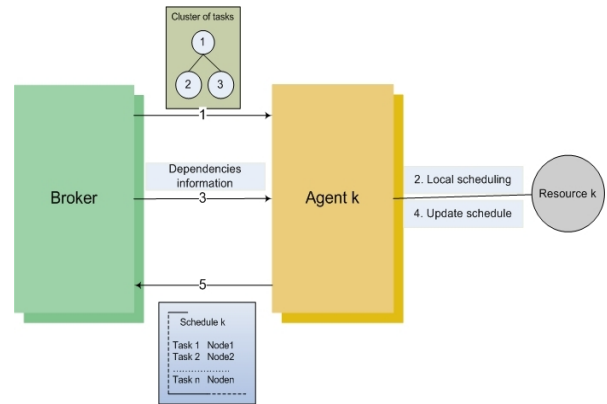

**Figure 5. Protocol 3**

# 4. Experimental results

This chapter presents the test scenarios, the evaluation criteria and the experimental results which enable the performance assessment of the CoAllocation strategy.

## 4.1. Test scenarios

The scenarios used for testing the scheduling algorithm are designed in order to observe and asses the performance indicators of the strategy.
A test scenario is described by the following aspects:
*Architecture:* specifies the resources used for testing.
*Actors:* describes the entities involved in the process.
*Input:* specifies the input file which contains the tasks which must be scheduled.
*The results* a test provides are presented in a graphic manner and are useful for the monitoring of the performance indicators of the algorithm. Every scheduling algorithm can be assessed by specific criteria.
*The performance indicators* are defined in order to obtain a valid assessment of the efficiency properties of CoAllocation mechanism. The indicator used by the performed tests, is the *number of tasks per agent*, which represents the number of tasks scheduled by each agent.

## 4.2. Test Architecture

All the tests were performed on the same architecture which consists of 6 workstations; these stations are part of a cluster, called *MinervaCluster*. Three agents run on these stations and manage them as shown in figure 6:

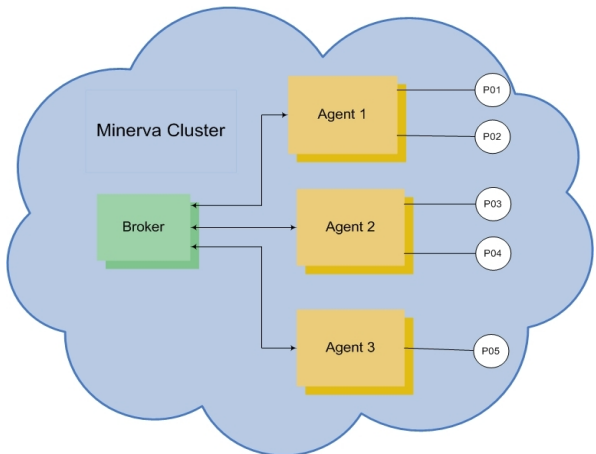

**Figure 6. Test architecture**

### 4.3. A test example

The DAG received as input, describing the set of tasks with dependencies is shown in figure 7:

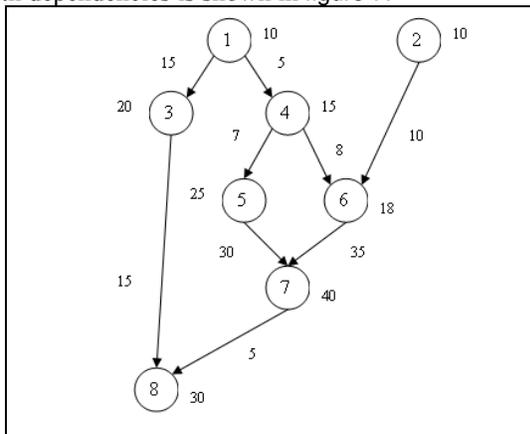

**Figure 7. The input DAG**

The obtained results are shown in figures 8 and 9.

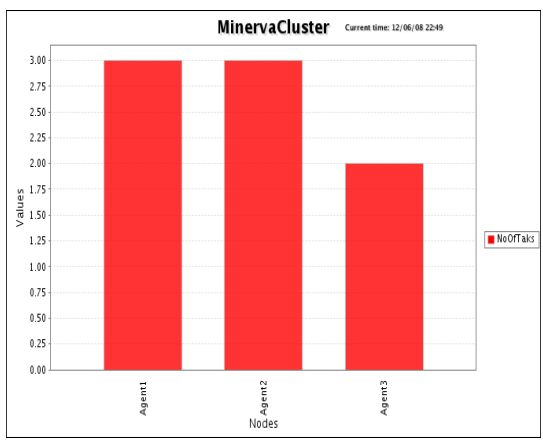

**Figure 8. Test1 – No. of tasks**

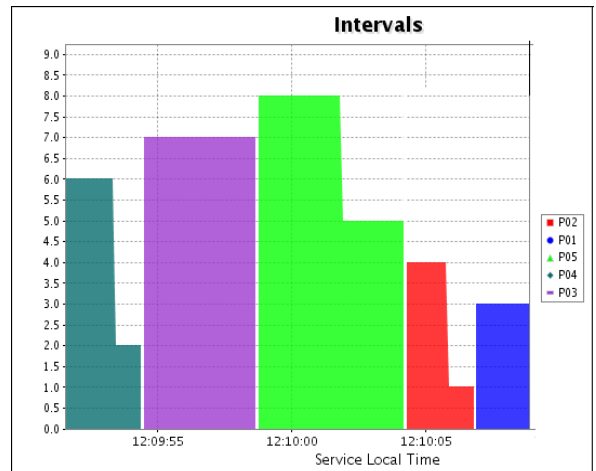

**Figure 9. Task Scheduling**

On the OY axis are shown the task Ids and the OX axis is a scaled representation of the intervals in which the tasks are scheduled.

As figures 8 and 9 show, the number of tasks is equally divided by the broker, so that:

- agent 1 (resources P01 and P02) schedules tasks 1, 3 and 4
- agent 2 (resources P03 and P04) receives tasks 2, 6 and 7
- agent 3 (resource P05) schedules tasks 5 and 8.

### 4.4. Performance evaluation

The tests presented in the previous sections, show that the algorithm schedules the input tasks by achieving a load balancing in terms of number of tasks distributed to each agent.

The graph in figure 10 summarizes the number of tasks per agent from all three tests.

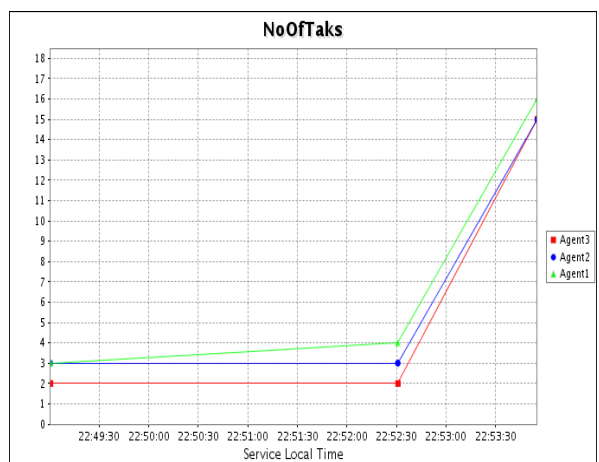

**Figure 10. No. of tasks**

## 4.5. Additional Tools

In order to obtain real-time monitoring and visualization of the performed tests, two existing tools were used.

The *MonALISA* [5] framework enables applications to log information that can be then visualized and analyzed in various ways. In order to achieve reliability and data consistency, MonALISA uses a secure and scalable communication.

*ApMon* [5] is a tool that can be used by an application to send monitoring data to MonALISA services. The data is sent as UDP datagrams to a specified MonALISA service; the use of the UDP protocol allows a great number of datagrams per second to be exchanged between ApMon and a running farm.

## 5. Related work

*Koala* [6] is a Grid scheduler that provides support for co-allocation. In the job model Koala uses, a job contains several components, which can run concurrently, at the same time and for the same amount of time. Koala uses local resource managers to which it distributes job components; the components of a single job may be assigned to several local resource managers. The specification of a component can include some restriction regarding the local resource manager, the number of resources to use, etc. The resources a job allocates are reserved for the entire execution time of the job.

Koala's architecture contains runners, which have several functions:

- ❍ Translation of user requests into a generic request language;
- ❍ Reservation, claiming and release of allocated resources;

The components of an application are started at the same time on different resources; prior to that, the input files the components need are being transferred to the corresponding location.

The algorithm uses the *Close to File* policy in order to minimize the transfer time of the necessary files to the resource allocated to a component. The *Close to File* policy places the tasks on resources that are closer to the storage location of the files a task requires.

The load balancing of resources is achieved by using the *Worst Fit* policy that places a job component on the resource with the largest number of idle processors.

The phases of the algorithm are:

- Place job components
- Transfer necessary input files

- Claim allocated processor
- Start components execution

The algorithm this paper proposes and the one Koala uses, they both have as purpose obtaining a load balancing of resources. The algorithm this paper proposes divides the tasks received as input into clusters, so that the agents can run the scheduling algorithm at the same time, minimizing this way, the duration of the algorithm.

## 6. Conclusions

Scheduling applications on wide-area distributed systems is useful for obtaining quick and reliable results in an efficient manner. Optimized scheduling algorithms are fundamentally important in order to achieve optimized resources utilization [7]. The existing and potential applications include many fields of activity like satellite image processing and medicine.

This paper proposes a scheduling algorithm for tasks with dependencies in Grid environments. The previous chapters have shown some basic aspects of the proposed strategy:

- The format of the input file (that contains the tasks description) and the description of a resource need to be in compliance with a standard format;
- The dependencies between the tasks (the algorithm treats only data dependencies) are best represented by a DAG; this association allowed the development of an efficient algorithm for coallocating the nodes of a DAG;
- The algorithm involves three phases and two types of entities, broker and agent, that are both involved in the scheduling algorithm;
- The purpose of the algorithm is to obtain load balancing among all the resources of the system, in terms of number of tasks scheduled on each resource;
- The tests and the graphical representations of the obtained results have shown that the strategy achieved its goal;

The algorithm described in this paper can be further extended to include more restrictions, such as tasks requirements, maximum communication cost inside a cluster, etc.